# Post-Treatment Confounding in Causal Mediation Studies:
# A Cutting-Edge Problem and A Novel Solution via Sensitivity Analysis

Guanglei Hong*[1], Fan Yang*[2], Xu Qin[3]
(* Equal first-authors)

[1] University of Chicago; Postal Address: 1126 E 59th Street, Chicago, IL 60637; Email: ghong@uchicago.edu
[2] University of Colorado Denver; Postal Address: 13001 E. 17th Place, Aurora, CO 80045; Email: fan.3.yang@cuanschutz.edu
[3] University of Pittsburgh; Postal Address: 5500 Wesley W. Posvar Hall, 230 S Bouquet St, Pittsburgh, PA 15260; Email: xuqin@pitt.edu

**Abstract:**

In causal mediation studies that decompose an average treatment effect into a natural indirect effect (NIE) and a natural direct effect (NDE), examples of post-treatment confounding are abundant. Past research has generally considered it infeasible to adjust for a post-treatment confounder of the mediator-outcome relationship due to incomplete information: it is observed under the actual treatment condition while missing under the counterfactual treatment condition. This study proposes a new sensitivity analysis strategy for handling post-treatment confounding and incorporates it into weighting-based causal mediation analysis without making extra identification assumptions. Under the sequential ignorability of the treatment assignment and of the mediator, we obtain the conditional distribution of the post-treatment confounder under the counterfactual treatment as a function of not just pretreatment covariates but also its counterpart under the actual treatment. The sensitivity analysis then generates a bound for the NIE and that for the NDE over a plausible range of the conditional correlation between the post-treatment confounder under the actual and that under the counterfactual conditions. Implemented through either imputation or integration, the strategy is suitable for binary as well as continuous measures of post-treatment confounders. Simulation results demonstrate major strengths and potential limitations of this new solution. A re-analysis of the National Evaluation of Welfare-to-Work Strategies (NEWWS) Riverside data reveals that the initial analytic results are sensitive to omitted post-treatment confounding.

A causal mediation analysis decomposes the average treatment effect on an outcome (ATE) into a natural indirect effect (NIE) and a natural direct effect (NDE). Past research has generally considered it infeasible to adjust for a post-treatment confounder when there exists treatment-by-mediator interaction (Avin, Shpitser, & Pearl, 2005; Robins, 2003). A post-treatment confounder may be viewed as an additional mediator that precedes the focal mediator on causal pathways. Handling post-treatment confounding is a cutting-edge problem in causal mediation analysis. This is because a post-treatment confounder is only *partially observed*. Specifically, if an individual has been assigned to the experimental condition, the individual's potential post-treatment confounder value associated with the counterfactual control condition is unobserved. In this sense, the problem with statistical adjustment for a post-treatment confounder is *a problem of missing data*.

This study proposes a new sensitivity analysis strategy for handling post-treatment confounding and incorporates it into weighting-based causal mediation studies when the treatment is randomized. The key is to obtain, for individuals in the experimental group, the conditional distribution of the post-treatment confounder under the counterfactual control condition, which is then adjusted for along with the distribution of the same confounder under the actual experimental condition. The analyst can make the adjustment flexibly through ratio-of-mediator-probability weighting (RMPW) (Hong, 2010, 2015; Hong & Nomi, 2012; Hong, Deutsch, & Hill, 2015; Lange, Vansteelandt, & Bekaert, 2012; Tchetgen Tchetgen & Shpister, 2012). Corresponding to a plausible range of the conditional correlation between the post-treatment confounder under the experimental condition and that under the control condition, our analytic procedure generates a bound for the NIE and that for the NDE, and thereby enabling the analyst to assess the sensitivity of the initial results to the omitted confounding.

Researchers have proposed several alternative strategies for handling post-treatment confounding in the presence of treatment-by-mediator interaction. The first is to change the causal estimands such that the research interest is no longer in decomposing the ATE into NIE and NDE (Geneletti, 2007; Rudolph et al, 2018; VanderWeele, Vansteelandt, & Robins, 2014; Wodtke & Zhou, 2020) except under special conditions (Vansteelandt & Daniel, 2017). The second is to invoke an extra assumption about the ignorability of $Z$ along with a series of model-based assumptions within the linear structural model framework (Imai & Yamamoto, 2013; Daniel et al, 2015). Past research that proposed to use weighting to adjust for post-treatment confounding similarly invoked additional assumptions involving the post-treatment confounder (Hong, 2015; Hong, Qin, & Yang, 2018; Huber, 2014).

In contrast, our solution focuses on decomposing the ATE into an NIE and an NDE without invoking additional identification assumptions beyond the sequential ignorability of the treatment assignment and of the mediator. This article is organized as follows. Section I introduces the application study and illustrates the need for handling post-treatment confounding. Section II presents the theoretical rationale for our new solution. Section III lays out a sensitivity analysis strategy for assessing the potential consequence of omitting a post-treatment confounder measured on a continuum. Section IV extends the rationale to an omitted post-treatment binary confounder. Section V investigates the performance of the new method across a range of realistic scenarios through a series of simulations. Section VI demonstrates the implementation in the real-data application. Section VII concludes and discusses further extensions.

## I.  Application Context

The welfare-to-work reform bill introduced by the U.S. Congress in the mid-1990s was intended to wean welfare applicants off the cash assistance system by providing incentives for participation in the labor force. Shortly before the new legislation, a randomized evaluation was conducted for assessing the potential impacts of this radical overhaul of the welfare system.  In Riverside, California, 694 welfare applicants with young children between ages 3-5 were assigned at random to either an experimental condition ($T = 1$)—a labor force attachment (LFA)

program—or a control condition ($T = 0$). The LFA program offered job search services and incentives including a threat of sanctions should one fail to meet the program requirements for actively seeking and securing employment; in contrast, the control group members were guaranteed cash assistance without the requirement for employment. Welfare mothers' psychological well-being was of potential concern because many were single mothers already disproportionately depressed at the baseline. Under the experimental condition in particular, the prospect of potentially losing the public safety net would likely trigger or aggravate depression.

**Research Questions and Causal Estimands**

Hong, Deutsch, and Hill (2015) investigated whether a treatment-induced change in employment played a mediating role in transmitting the program impact on maternal depression ($Y$) two years later. The mediator ($M$) indicated one's post-treatment employment record prior to the measure of depression. For simplicity, here we re-examine the results based on a binary mediator that takes values $M = 1$ if an individual was ever employed and $M = 0$ if the individual was never employed during the two years after randomization. The following research questions correspond to the ATE, the NIE, and the NDE:

*ATE*: What is the average impact of the LFA program on maternal depression 2 years later?

*NIE*: How much of this impact is attributable to program-induced change in employment?

*NDE*: What would be the average program impact on maternal depression if the program failed to change employment status?

Each causal effect is defined in terms of potential outcomes:

$$ATE = E[Y(1) - Y(0)] \equiv E\big[Y\big(1, M(1)\big) - Y\big(0, M(0)\big)\big];$$

$$NIE = E\big[Y\big(1, M(1)\big) - Y\big(1, M(0)\big)\big];$$

$$NDE = E\big[Y\big(1, M(0)\big) - Y\big(0, M(0)\big)\big].$$

The sum of NIE and NDE is equal to ATE (Pearl, 2001; Robins & Greenland, 1992). Here $M(1)$ and $M(0)$ denote an individual's potential employment status under the experimental condition and the control condition, respectively; $Y(1, M(1))$ and $Y(0, M(0))$ denote the individual's potential depression level under the experimental condition and the control condition, respectively; the third potential outcome $Y(1, M(0))$ denotes the same individual's potential depression level under the experimental condition should the treatment counterfactually fail to change the individual's employment status from its value associated with the control condition $M(0)$ to that associated with the experimental condition $M(1)$. Each potential mediator or potential outcome is a random variable that "naturally" takes different values. The NIE is the average impact of the program on maternal depression associated with a change from $M(0)$ to $M(1)$ in a hypothetical scenario in which all individuals would be subjected to the new policy requirements; the NDE is the program impact on maternal depression in a hypothetical scenario in which all individuals' employment status would remain unchanged by the program.

**Initial Results and Potential Bias**

Although individuals were assigned at random to one of the two treatment conditions, they were not assigned at random to different mediator values under each treatment condition. Researchers of the original study (Hong et al, 2015) made adjustment for a set of pretreatment covariates including demographics, family structure, educational attainment, baseline measures of depressive symptoms, prior employment history, and prior history of welfare dependence through an RMPW analysis. This adjustment strategy simply transforms through weighting the mediator distribution of the experimental group to resemble that of the control group within levels of the pretreatment covariates. The average weighted outcome of the experimental group identifies the average counterfactual outcome $E[Y(1, M(0))]$ under the assumption of no omitted confounders. Importantly, the RMPW strategy allows for treatment-by-mediator interaction. This is relevant in the current application in which there was clear evidence that the mediator-outcome relationship differed between the two treatment conditions. As anticipated, an individual's failure in seeking employment seemingly heightened depressive symptoms under the experimental condition but not under the control condition. More generally, RMPW is flexible for handling any types of nonlinearity because it does not require an explicit specification of the outcome model; it also has the flexibility of accommodating multi-category and multi-valued mediators.

The researchers reported tentative evidence that, on the one hand, assignment to LFA indeed increased employment rate from 40% to 65%, which would lead to a considerable reduction in depression on average as a result; on the other hand, should the employment rate fail to improve, assignment to LFA would increase depression on average. However, neither estimate reached the conventional statistical significance level for rejecting the null hypothesis. The identification required the *sequential ignorability* assumption (Imai, Keele, & Yamamoto, 2010):

(1) *Ignorable treatment assignment* given the observed pretreatment covariates $\mathbf{X} = \mathbf{x}$:

$$Y(t, m), M(t), M(t') \perp\!\!\!\perp T \mid \mathbf{X} = \mathbf{x};$$

(2) *Ignorable mediator value assignment* under each treatment condition given $\mathbf{X} = \mathbf{x}$:

$$Y(t, m) \perp\!\!\!\perp M(t), M(t') \mid T = t, \mathbf{X} = \mathbf{x}.$$

The sequential ignorability assumption would be violated in the presence of omitted pretreatment confounders. Hong, Qin, & Yang (2018) proposed a weighting-based approach to sensitivity analysis that is integrated with the RMPW strategy for causal mediation analysis. Conducting a sensitivity analysis with the same data, they reported that the initial estimate of NIE was sensitive to bias associated with some of the omitted pretreatment covariates.

The sequential ignorability assumption would also be violated in the presence of a post-treatment covariate that confounds the mediator-outcome relationship under either or both treatment conditions. One example is post-treatment welfare amount. On average, individuals in the experimental group received significantly less welfare amount than did their counterparts in the control group during the first year after randomization. A higher amount of post-treatment welfare appears to be associated with a lower propensity of employment. Another example is whether an individual was continuously on welfare during the first year after randomization. The proportion of such individuals was significantly higher in the control group than in the experimental group. These individuals were also less likely to become employed.

We use $Z$ to denote a post-treatment covariate. To clarify the unique challenge to causal mediation analysis in the presence of post-treatment confounding despite the randomization of treatment assignment, we illustrate with a pair of graphs each representing one of the two treatment conditions. As shown in Figure 1a, individual $i$ under the control condition would display potential post-treatment covariate $Z_i(0)$, potential mediator $M_i(0)$, and potential outcome $Y_i(0)$. Figure 1b shows that the same individual, under the alternative experimental condition, would display potential post-treatment covariate $Z_i(1)$, potential mediator $M_i(1)$, and potential outcome $Y_i(1)$. Under treatment condition $t$ for $t = 0,1$, a vector of pretreatment covariates $\mathbf{X}_i$ may predict $Z_i(t)$, $M_i(t)$, and $Y_i(t)$. Moreover, a vector of unobserved pretreatment covariates $\mathbf{U}_i$ along with a set of individual-specific random events denoted by $\epsilon_i$ may predict $Z_i(t)$ and $Y_i(t)$. In the current application, examples of $\mathbf{U}$ might include whether an individual was financially dependent on relatives at the baseline; examples of $\epsilon$ might include whether important communications from the social services administration office were accidentally lost in mail which could cause an unintended lapse of welfare. We relax the second component of the sequential ignorability assumption by instead assuming ignorable mediator value assignment given $\mathbf{X}_i$ and $Z_i(t)$. Under this assumption, $\mathbf{U}_i$ and $\epsilon_i$ are conditionally independent of $M_i(1)$ and $M_i(0)$; under the same assumption, there is no cross-world conditional association between $M_i(1)$ and $Y_i(0)$ and between $M_i(0)$ and $Y_i(1)$.

However, it is well-known that statistical adjustment for a post-treatment covariate would introduce bias in identifying the ATE (Rosenbaum, 1984). The observed value of $Z_i$ is related to the potential values of the post-treatment covariate as follows: $Z_i = T_i Z_i(1) + (1 - T_i) Z_i(0)$. Because $Z(1)$ and $Z(0)$ are not expected to be equal in the population, comparing individuals in the experimental group whose $Z(1) = z$ with those in the control group whose $Z(0) = z$ is tantamount to comparing "apples" and "oranges" in many cases. To remove bias associated with the post-treatment confounder in identifying the NIE and NDE, it is necessary to make statistical adjustment for both $Z(0)$ and $Z(1)$ along with **X**. The fundamental difficulty is that the analyst could observe either $Z(0)$ or $Z(1)$ but not both.

## II. Theoretical Rationale for the New Solution

In the presence of a post-treatment confounder that precedes the focal mediator, the key challenge is to use the observed information to identify $E\big[Y\big(1, M(0)\big)\big]$. We modify the sequential ignorability assumptions as follows for $t, t' = 0, 1$ and $t \neq t'$:

(i)    *Ignorable treatment assignment* given the observed pretreatment covariates:

$$Y(t,m), M(t), M(t'), Z(t), Z(t') \perp\!\!\!\perp T \,|\, \mathbf{X} = \mathbf{x}.$$

Given that $Z(t)$ and $Z(t')$ are each specified under a respective fixed treatment condition, the following result can be easily derived from assumption (i). This variant of assumption (i) states that the treatment assignment is ignorable given the potential post-treatment covariate values $Z(t) = z$ and $Z(t') = z'$ in addition to $\mathbf{X} = \mathbf{x}$. Appendix 1 shows the derivation.

$$Y(t,m), M(t), M(t') \perp\!\!\!\perp T \,|\, Z(t) = z, Z(t') = z', \mathbf{X} = \mathbf{x}.$$

Assumption (i) and its variant are guaranteed when the treatment is randomized possibly conditional on **X**.

(ii)    *Ignorable mediator value assignment* under each treatment condition given the observed pretreatment and post-treatment covariates:

$$Y(t,m), Y(t',m) \perp\!\!\!\perp M(t) \,|\, T = t, Z(t) = z, \mathbf{X} = \mathbf{x};$$

$$Y(t,m) \perp\!\!\!\perp M(t') \,|\, T = t, Z(t) = z, \mathbf{X} = \mathbf{x}.$$

Assumption (ii) is considerably more plausible than the standard ignorability assumption (2) discussed earlier because post-treatment confounding is often inevitable. Under assumption (ii), when $Z(t)$ and **X** are given for individuals assigned to treatment condition $t$, the relationship between $Y(t,m)$ and $M(t)$, that between $Y(t,m)$ and $M(t')$, and that between $Y(t',m)$ and $M(t)$ will not be confounded by $Z(t')$ or by any other pretreatment or post-treatment covariates. This assumption rules out cross-world connections between the mediator and the outcome, which also implies that there is no cross-world connection between $M(t)$ and $Z(t')$:

$$M(t) \perp\!\!\!\perp Z(t') \mid T = t, Z(t) = z, \mathbf{X} = \mathbf{x}.$$

This variant of assumption (ii) can be understood as follows: $Y(t', m)$ is a function of $Z(t')$ conditioning on $T = t, Z(t) = z$ and $\mathbf{X} = \mathbf{x}$. If $M(t)$ and $Z(t')$ were not conditionally independent, $M(t)$ would have a back-door path to $Y(t', m)$ via $Z(t')$, which would violate assumption (ii). The following theoretical results are key to our new solution.

**THEOREM**.

Under assumptions (i) and (ii),

$$P(M(0) = m \mid T = 1, Z(1) = z, \mathbf{X} = \mathbf{x})$$

$$= \int P(M(0) = m \mid T = 0, Z(0) = z', \mathbf{X} = \mathbf{x}) h_{z'} dz',$$

where

$$h_{z'} = h_{z'}(\mathbf{x}) = P(Z(0) = z' \mid T = 1, Z(1) = z, \mathbf{X} = \mathbf{x}).$$

See Appendix 2 for the proof. This theorem states that, for an individual who has been assigned to $T = 1$ and displays covariate values $\mathbf{X} = \mathbf{x}$ and $Z(1) = z$, the individual's conditional probability of displaying a certain mediator value under the counterfactual control condition $M(0) = m$ can be obtained when we relate it to the conditional probability of the mediator for the individual's counterparts who have actually been assigned to $T = 0$ with covariate values $\mathbf{X} = \mathbf{x}$ and $Z(0) = z'$. Importantly, for the focal individual who has been assigned to $T = 1$, the conditional probability of $M(0)$ under the counterfactual condition needs to be averaged over the individual's conditional distribution of $Z(0)$ denoted by $h_{z'}$.

**LEMMA**.

Under assumptions (i) and (ii) and when the treatment is randomized, NIE and NDE can be identified through weighting:

$$NIE = E[Y|T = 1] - E\left[\int W(z')Yh_{z'}dz' \,|T = 1\right];$$

$$NDE = E\left[\int W(z')Yh_{z'}dz' \,|T = 1\right] - E[Y|T = 0];$$

$$W(z') = \frac{P(M(0) = m|T = 0, Z(0) = z', \mathbf{X = x})}{P(M(1) = m|T = 1, Z(1) = z, \mathbf{X = x})}.$$

Appendix 3 provides a proof of this Lemma. It is easy to extend the weight to the case in which the treatment assignment is random within levels of $\mathbf{X = x}$.

This theoretical result enables sensitivity analysis for assessing the potential consequence of omitting a post-treatment confounder. In the subsequent sections, we propose and evaluate analytic strategies for empirically estimating the parameters that define the distribution of $Z(0)$ when $\mathbf{X = x}$ and $Z(1) = z$ are given. The conditional distribution of $Z(0)$ depends on the unknown conditional correlation between $Z(0)$ and $Z(1)$ regarded as a sensitivity parameter.

### III. Sensitivity Analysis for a Post-Treatment Continuous Confounder

For a post-treatment confounder measured on a continuum, assuming a bivariate normal distribution of $Z(0)$ and $Z(1)$ conditioning on the observed pretreatment covariates $\mathbf{X = x}$, we obtain the conditional distribution of $Z(0)$ for individuals assigned to $T = 1$. To implement, one strategy is to impute $Z(0)$ by taking multiple random draws from its conditional distribution; an alternative strategy is to take the integral of $Z(0)$ over its conditional distribution.

**Theoretical Models**

As illustrated in Figures 1a and 1b, $Z_i(0)$ and $Z_i(1)$ are each a function of observed pretreatment covariates $\mathbf{X}_i$, unobserved pretreatment covariates $\mathbf{U}_i$, and the impact of random events $\epsilon_i$. In general, the causal effect of $T$ on $Z$ may depend on $\mathbf{X}$, $\mathbf{U}$, and $\epsilon$. Because the impacts of $\mathbf{U}_i$ and $\epsilon_i$ on $Z_i(t)$ are not empirically distinguishable, we use $r_{ti}$ to denote their joint impact:

$$Z_i(0) = \mu(0, \mathbf{x}) + r_{0i};$$

$$Z_i(1) = \mu(1, \mathbf{x}) + r_{1i}.$$

Let $\mu(t, \mathbf{x})$ denote a function of $\mathbf{x}$ under treatment condition $t$ for $t = 0, 1$; $\mu(t, \mathbf{x})$ may take any flexible parametric or nonparametric functional form. For individual $i$ whose $T_i = 1, Z_i(1) = z$, and $\mathbf{X}_i = \mathbf{x}$, due to the randomness of $r_{1i}$ and $r_{0i}$, the counterfactual $Z_i(0)$ is a random variable. To obtain the conditional distribution of $Z_i(0)$ given $Z_i(1)$ and $\mathbf{X}_i$, we assume that:

$$\begin{pmatrix} r_{0i} \\ r_{1i} \end{pmatrix} \sim N\left( \begin{pmatrix} 0 \\ 0 \end{pmatrix}, \begin{pmatrix} \sigma_0^2 & \rho_{10}\sigma_1\sigma_0 \\ \rho_{10}\sigma_1\sigma_0 & \sigma_1^2 \end{pmatrix} \right).$$

Here $r_{1i}$ and $r_{0i}$ are assumed to be bivariate normal; $\sigma_t^2 = var(r_{ti}|T_i = t, \mathbf{X}_i = \mathbf{x})$ for $t = 0, 1$; $\rho_{10} = corr(Z(1), Z(0)|T_i = t, \mathbf{X} = \mathbf{x}) = corr(Z(1), Z(0)|\mathbf{X} = \mathbf{x})$ under assumption (i). By convention, $\sigma_1$, $\sigma_0$, and $\rho_{10}$ are assumed to be invariant across different values of $\mathbf{x}$. We will investigate potential implications of these distributional assumptions through simulations presented in section VI.

Under the above specification, we obtain the conditional distribution of $Z_i(0)$ for individual $i$ whose $T_i = 1, Z_i(1) = z$, and $\mathbf{X}_i = \mathbf{x}$:

$$E[Z_i(0)|T_i = 1, Z_i(1) = z, \mathbf{X}_i = \mathbf{x}] = \mu(0, \mathbf{x}) + \rho_{10}\frac{\sigma_0}{\sigma_1}\big(z - \mu(1, \mathbf{x})\big);$$

$$var(Z_i(0)|T_i = 1, Z_i(1) = z, \mathbf{X}_i = \mathbf{x}) = (1 - \rho_{10}^2)\sigma_0^2. \tag{1}$$

In its essence, $\rho_{10}$ determines the relative contribution of the observed value of $Z_i(1)$ in predicting the counterfactual $Z_i(0)$ for individual $i$. The value of $\rho_{10}$, which is bounded between -1 and 1, cannot be empirically obtained. To further narrow the bounds, the analyst may utilize an additional baseline covariate $C$. Let $\rho_{1C} = corr(Z, C|T = 1, \mathbf{X} = \mathbf{x})$ and $\rho_{0C} = corr(Z, C|T = 0, \mathbf{X} = \mathbf{x})$ denote the partial correlations. As shown in past research (Olkin, 1981; Stangley & Wang, 1969; also see Yang et al., 2017), $\rho_{10}$ is restricted by the inequalities:

$$\rho_{1C}\rho_{0C} - \sqrt{(1 - \rho_{1C}^2)(1 - \rho_{0C}^2)} \le \rho_{10} \le \rho_{1C}\rho_{0C} + \sqrt{(1 - \rho_{1C}^2)(1 - \rho_{0C}^2)}. \tag{2}$$

When $J$ additional covariates are available, the analyst may compute $J$ bounded sets, the intersection of which defines a conservative bounded set of values for $\rho_{10}$.

**Analytic Steps**

There are four major steps in conducting the sensitivity analysis. Step 3 can be carried out through either imputation or integration. They are described as Step 3a and Step 3b, respectively.

*Step 1. Predict $Z$ as a function of observed baseline covariates and obtain the residuals.* For individuals with $Z = z$ and $\mathbf{X} = \mathbf{x}$, the residuals are $\hat{r}_0 = z - \widehat{\mu(0, \mathbf{x})}$ if $T = 0$ and $\hat{r}_1 = z - \widehat{\mu(1, \mathbf{x})}$ if $T = 1$; the respective sample variances are $\widehat{\sigma_0^2}$ and $\widehat{\sigma_1^2}$. Importantly, the set of baseline covariates that predict $Z$ should contain but should not be limited to the set of covariates that confound the mediator-outcome relationship; and the prediction model for $Z$ may take any flexible form. This is because, as we will show through the simulations, including strong predictors of Z may reduce the range of the bounds in a sensitivity analysis.

*Step 2. Obtain the conditional distribution of $Z(0)$.* The analyst may choose a set of evenly spaced hypothetical values of $\rho_{10}$ within its bounds including the values at the upper and lower bounds. Corresponding to a given hypothetical value of $\rho_{10}$, the parameters of the conditional distribution of $Z_i(0)$ are estimated for individual $i$ assigned to $T_i = 1$ when $\mu(0, \mathbf{x})$, $\mu(1, \mathbf{x})$, $\sigma_0$, and $\sigma_1$ in equation (1) are replaced by their sample analogues.

*Step 3a. Use imputed values of $Z(0)$ in an RMPW analysis.* When implementing the Lemma through multiple imputation (Little & Rubin, 2019), the analyst may take $K$ random draws from the conditional distribution of $Z(0)$. Let $z'_{ik}$ be the $k$th random draw for individual $i$ with $T_i = 1$, $Z_i(1) = z$, $\mathbf{X}_i = \mathbf{x}$, and $M_i(1) = m$. The imputed values are viewed as given in each set of imputed data. One may then estimate the ratio-of-mediator-probability weight:

$$\widehat{W_{ik}} = \frac{\hat{P}(M_i(0) = m | T_i = 0, Z_i(0) = z'_{ik}, \mathbf{X}_i = \mathbf{x})}{\hat{P}(M_i(1) = m | T_i = 1, Z_i(1) = z, \mathbf{X}_i = \mathbf{x})}.$$

To implement, a separate propensity score model for the mediator is fitted to the data in each of the two treatment groups. The denominator is individual $i$'s estimated propensity of displaying the observed mediator value $m$ under the experimental condition. The model fitted to the control group data is used for predicting the same individual's propensity of displaying the same mediator value under the control condition as a function of $\mathbf{x}$ and the imputed value $z'_{ik}$. The sample estimator of NIE and that of NDE are averaged over the $K$ random draws of $Z(0)$:

$$\widehat{NIE} = \frac{1}{K}\sum_{k=1}^{K}\left[\frac{\sum_i T_i Y_i}{\sum_i T_i} - \frac{\sum_i T_i \widehat{W_{ik}} Y_i}{\sum_i T_i \widehat{W_{ik}}}\right];$$

$$\widehat{NDE} = \frac{1}{K}\sum_{k=1}^{K}\left[\frac{\sum_i T_i \widehat{W_{ik}} Y_i}{\sum_i T_i \widehat{W_{ik}}} - \frac{\sum_i (1-T_i) Y_i}{\sum_i (1-T_i)}\right].$$

In accordance with Rubin's rules (Little & Rubin, 2019), the standard error for $\widehat{NIE}$ and that for $\widehat{NDE}$ are each pooled over the $K$ estimates.

*Step 3b. Integrate over the predicted distribution of $Z(0)$ in an RMPW analysis.* Alternatively, we may implement the Lemma through taking the integral with respect to $z'$ over the conditional distribution of $Z_i(0)$ for individual $i$ whose $T_i = 1, Z_i(1) = z, \mathbf{X}_i = \mathbf{x}$, and $M_i(1) = m$. We estimate $W_i(z')$ as a function of $z'$, as specified in the Lemma, and apply the normal density function to $h_{iz'} = f(Z_i(0) = z'|T_i = 1, Z_i(1) = z, \mathbf{X}_i = \mathbf{x})$ in the integration. Subsequently, the sample estimator of NIE and that of NDE can be obtained as follows:

$$\widehat{NIE} = \frac{\sum_i T_i Y_i}{\sum_i T_i} - \frac{\sum_i T_i Y_i \int \widehat{W_i(z')}\widehat{h_{iz'}} dz'}{\sum_i T_i \int \widehat{W_i(z')}\widehat{h_{iz'}} dz'};$$

$$\widehat{NDE} = \frac{\sum_i T_i Y_i \int \widehat{W_i(z')}\widehat{h_{iz'}} dz'}{\sum_i T_i \int \widehat{W_i(z')}\widehat{h_{iz'}} dz'} - \frac{\sum_i (1-T_i) Y_i}{\sum_i (1-T_i)}.$$

Extending the previous results for RMPW-based causal mediation analysis that takes into account the estimation uncertainty in the propensity score-based weight (Bein et al, 2018), we derive the asymptotic standard error for $\widehat{NIE}$ and that for $\widehat{NDE}$ when the integration method is employed and obtain estimates of these standard errors accordingly.

*Step 4. Sensitivity analysis.* Steps 2 and 3 are repeated at each hypothetical value of $\rho_{10}$. Corresponding to the bounds for $\rho_{10}$ are the bounds for NIE and NDE estimates. These are to be contrasted with the initial estimates of NIE and NDE, respectively. The analyst may determine that the initial results are sensitive to the omission of the post-treatment confounder if they clearly deviate from the bounds for NIE and NDE estimates obtained from the current analysis. In addition, to assess whether the results of hypothesis testing are sensitive to the omission of post-treatment confounding, the analyst may estimate the confidence bands for NIE and NDE across the range of $\rho_{10}$. These are to be contrasted with the respective confidence intervals for NIE and NDE obtained from the initial analysis.

### IV. Sensitivity Analysis for a Binary Post-Treatment Confounder

We now present our strategy for obtaining the conditional distribution of $Z_i(0)$ for individual $i$ in the experimental group when the post-treatment confounder is binary.

**Theoretical Models**

We adopt a bivariate probit model for $Z_i(0)$ and $Z_i(1)$:

$$Z_i^*(0) = \mu(0, \mathbf{x}) + r_{0i}, \qquad Z_i(0) = I(Z_i^*(0) > 0);$$

$$Z_i^*(1) = \mu(1, \mathbf{x}) + r_{1i}, \qquad Z_i(1) = I(Z_i^*(1) > 0);$$

$$\begin{pmatrix} r_{0i} \\ r_{1i} \end{pmatrix} \sim N\left( \begin{pmatrix} 0 \\ 0 \end{pmatrix}, \begin{pmatrix} 1 & \rho_{10} \\ \rho_{10} & 1 \end{pmatrix} \right).$$

Here $I(\cdot)$ denotes an indicator function; $Z_i(t)$ takes value 1 if and only if its corresponding continuous latent variable $Z_i^*(t)$ is positive; $\mu(t, \mathbf{x})$ may take any flexible parametric or nonparametric functional form; $\rho_{10} = corr(Z^*(1), Z^*(0)|\mathbf{X} = \mathbf{x})$. Under the above specification, we have that

$$P(Z_i(0) = 1|T_i = 1, Z_i(1) = 1, \mathbf{X}_i = \mathbf{x}) = P\big(r_{0i} > -\mu(0, \mathbf{x})\big|r_{1i} > -\mu(1, \mathbf{x})\big);$$

$$P(Z_i(0) = 1|T_i = 1, Z_i(1) = 0, \mathbf{X}_i = \mathbf{x}) = P\big(r_{0i} > -\mu(0, \mathbf{x})\big|r_{1i} \leq -\mu(1, \mathbf{x})\big).$$

Let $\Phi(\cdot)$ denote the cumulative distribution function (CDF) of a standard normal random variable; and let $\boldsymbol{\Phi}_\rho(\cdot, \cdot)$ denote the CDF of bivariate standard normal random variables with correlation $\rho$. The conditional probability of $Z_i(0)$ can be represented as follows:

$$P(Z_i(0) = 1|T_i = 1, Z_i(1) = 1, \mathbf{X}_i = \mathbf{x}) = 1 - \frac{\Phi(-\mu(0,\mathbf{x})) - \mathbf{\Phi}_{\rho_{10}}(-\mu(0,\mathbf{x}), -\mu(1,\mathbf{x}))}{\Phi(\mu(1,\mathbf{x}))},$$

$$P(Z_i(0) = 1|T_i = 1, Z_i(1) = 0, \mathbf{X}_i = \mathbf{x}) = 1 - \frac{\mathbf{\Phi}_{\rho_{10}}(-\mu(0,\mathbf{x}), -\mu(1,\mathbf{x}))}{\Phi(-\mu(1,\mathbf{x}))}.$$

Again, the analyst may constrain the bounds for $\rho_{10}$ by utilizing a baseline covariate $C$ that conditionally correlates with $Z(1)$ and $Z(0)$. For example, if $C$ is binary, let $\rho_{1C} = corr(Z^*(1), C^*|T = 1, \mathbf{X} = \mathbf{x})$ where $C^*$ is a latent continuous variable. The analyst may estimate this conditional correlation by fitting a bivariate probit model to the experimental group with $Z(1)$ and $C$ as response variables and $\mathbf{X}$ as explanatory variables. Similarly, let $\rho_{0C} = corr(Z^*(0), C^*|T = 0, \mathbf{X} = \mathbf{x})$, which can be estimated by fitting a bivariate probit model to the control group. $\rho_{10}$ is then restricted by the same inequalities as specified earlier in equation (2).

**Analytic Steps**

As before, there are four major steps in conducting the sensitivity analysis, in which Step 3 can be carried out through either imputation or integration. One difference is that, in step 1, the functions $\mu(0, \mathbf{x})$ and $\mu(1, \mathbf{x})$ are each estimated by fitting a univariate probit model to the respective treatment group. Another difference is that, in implementing the integration method in step 3b, we estimate the binomial density functions $h_{i1} = P(Z_i(0) = 1|T_i = 1, Z_i(1) = z, \mathbf{X}_i = \mathbf{x})$ and $h_{i0} = 1 - h_{i1}$. Subsequently, we obtain the estimates of NIE and NDE:

$$\widehat{NIE} = \frac{\sum_i T_i Y_i}{\sum_i T_i} - \frac{\sum_i T_i Y_i \sum_{z'=0}^1 \widehat{h_{iz'}} \; \widehat{W_i(z')}}{\sum_i T_i \sum_{z'=0}^1 \widehat{h_{iz'}} \; \widehat{W_i(z')}};$$

$$\widehat{NDE} = \frac{\sum_i T_i Y_i \sum_{z'=0}^1 \widehat{h_{iz'}} \; \widehat{W_i(z')}}{\sum_i T_i \sum_{z'=0}^1 \widehat{h_{iz'}} \; \widehat{W_i(z')}} - \frac{\sum_i (1 - T_i) Y_i}{\sum_i (1 - T_i)}.$$

**V. Simulations**

To assess the feasibility of this novel sensitivity analysis procedure and to evaluate its performance, the simulation study addresses two sets of research questions. The first set of questions concerns the performance of the imputation-based RMPW approach and the integration-based RMPW approach when the distributional assumptions about the post-treatment confounder are valid; the second set of questions is about their performance when such assumptions are invalid. Across 12 different scenarios of data generation, we use the oracle estimator of NIE and that of NDE as the benchmark to compare the results with. These oracle estimators are obtained by adjusting for the underlying true value of $Z_i(0)$ in the RMPW analysis instead of using the predicted distribution of $Z_i(0)$. The oracle estimators are consistent estimators of NIE and NDE. We also present the results for the naïve estimators of NIE and NDE when the RMPW analysis makes no adjustment for the post-treatment confounder.

In the imputation-based RMPW analysis, the number of imputations is set to be 25. In the integration-based RMPW analysis, we apply the method of Gauss-Hermite quadrature with 10 quadrature points to approximate the values of integrals when $Z$ is continuous. The sample size is $n = 2,000$ in most simulation scenarios except in the first set of simulations for making a contrast with a relatively small sample size $n = 200$; additionally, in simulation scenario 10, we consider a sample size similar to the application study with $n = 700$. The number of replications in each simulation scenario is 1,000. The supporting information provides details of the data generation. The simulation results for NIE and NDE are summarized in Tables 1 and 2, respectively.

**Performance When the Distributional Assumptions Are Valid**

*Research Question 1*: When the model for $Z$ is correctly specified, are the imputation-based RMPW estimator and the integration-based RMPW estimator consistent with their benchmark values when $\rho_{10}$ is equal to its true value?

There is clear evidence that, for both a continuous $Z$ and a binary $Z$, the results are consistent when $\rho_{10}$ is equal to its true value. This conclusion holds across a wide range of scenarios as shown in Table 1 for NIE and Table 2 for NDE. Specifically, simulation scenarios 1 to 6 evaluate the performance of the proposed estimators for a continuous $Z$; and scenario 11 for a binary $Z$. For a continuous $Z$, scenarios 1 to 3 consider a true value of $\rho_{10}$ being 0.5, 0, and - 0.5, respectively, when the sample size is 2,000; in parallel, scenarios 4 to 6 consider the same set of true values of $\rho_{10}$ when the sample size is reduced to 200.

   **Research Question 2**: For a correctly specified model for *Z*, does the result deviate from the benchmark value when the hypothetical value of $\rho_{10}$ deviates from its true value?

   As anticipated, the deviation from the benchmark value increases when the hypothetical value of $\rho_{10}$ is farther away from the true value. This is true for both a continuous *Z* and a binary *Z*. Figure 2 provides a graphical illustration based on simulation scenario 1 where *Z* is continuous with a true value of $\rho_{10}$ being 0.5 and with a sample size *n* = 2,000. For both the NIE and the NDE, the imputation-based estimate and the integration-based estimate match the benchmark values at the true value of $\rho_{10}$. These estimates increasingly deviate from the benchmark values as the hypothetical value of $\rho_{10}$ shifts farther away from its true value. The patterns are similar for different true values of $\rho_{10}$ (scenarios 2 and 3), for a binary *Z* (scenario 11), as well as for different sample sizes (scenarios 4 to 6).

   **Research Question 3**: Does the integration method outperform the imputation method in terms of the efficiency of estimation? Do the confidence bands accurately reflect the estimation error when the imputation method or the integration method is employed?

   When comparing the true standard error (i.e., the standard deviation of the NIE estimator or the NDE estimator over the 1,000 replications) between the imputation-based strategy and the integration-based strategy, we find no evidence that one is more efficient than the other. This is true for both a continuous *Z* and a binary *Z*. For both the imputation-based estimator and the integration-based estimator, there is a general tendency that the estimated standard error converges to the true standard error when the sample size increases. The estimated standard error for the imputation-based estimator tends to be slightly greater than the true standard error regardless of sample size. Comparing between the two different sample sizes (2,000 vs. 200), it appears that the imputation-based estimator is preferred when the sample size is relatively small as it tends to err on the conservative side; when the sample size is relatively large, the integration-based estimator is preferred as it tends to closely approximate the true standard error.

   ***Research Question 4***: When the prediction model for $Z$ does not match the data generation model, is the result still robust at its true value of $\rho_{10}$? Is there an increase in the width of the bounds for the estimated values of NIE and NDE over the range of plausible values of $\rho_{10}$? Is there an increase in the width of the confidence bands for NIE and NDE?

   When the prediction model for $Z$ specified by the analyst is different from the data generation model for $Z$, the former tends to explain less variation in $Z$ when compared with the latter; the true value of $\rho_{10}$ will differ as well between the prediction model and the data generation model due to the corresponding changes to the residuals $r_0$ and $r_1$. A comparison between scenarios 7a and 7b reveals the implications. In scenario 7a, the prediction model for $Z$ is a function of a single pretreatment covariate $X$, with $\rho_{10}$ = 0.8. In scenario 7b, the prediction model for $Z$ is the same as the underlying data generation model where $Z$ is a function of $X$ and an additional predictor denoted by $L$, with $\rho_{10}$ = 0.5. The variation in $Z(0)$ and that in $Z(1)$ explained by $X$ and $L$ are about five times the variations explained by $X$ alone. Figure 3 displays the estimation results. Despite the differences in how the structural part of the prediction model for $Z$ is specified, the point estimates of NIE and those of NDE are always consistent at the true values of $\rho_{10}$ corresponding to each model. However, the bounds for the NIE and NDE estimates become wider in scenario 7a where relatively less variation in $Z_i(0)$ and that in $Z_i(1)$ are explained. The impact on the width of the confidence bands appears to be negligible. We conclude that even though the structural part of the prediction model specified by the analyst tends to deviate from the data generation model, the estimation results are robust. Yet importantly, an increase in the predictive power of the model for $Z$ under each treatment condition will effectively reduce the width of the bounds for the NIE and NDE estimates.

**Performance When the Distributional Assumptions Are Violated**

**Research Question 5**: Is there a consequence when homoscedasticity is violated in the prediction model for a continuous $Z$?

In scenario 8, we specify $\sigma_0$, $\sigma_1$, and $\rho_{10}$ each to be a function of a pretreatment covariate; the analysis, however, assumes homoscedasticity. According to the results shown in Figure 4, violations of the homoscedasticity assumption do not appear to be consequential. At the value of the overall population correlation $\rho_{10}$ conditioning on the structural model, the NIE and NDE estimates are consistent with their respective benchmark values.

**Research Question 6**: Is there a consequence when multivariate normality is violated in the prediction model for $Z$?

We find that when violations of the multivariate normality assumption are not severe, such as in the case of a continuous $Z$ with a small degree of zero inflation (scenario 10) or in the case of a binary $Z$ generated with logistic random errors (scenario 12), the NIE and NDE estimates deviate only slightly from their respective benchmark values at the true value of $\rho_{10}$. However, when violations are severe, as in the case of a continuous $Z$ that follows a gamma distribution rather than a normal distribution (scenario 9), the estimation results are no longer robust. Figure 5 provides a graphical illustration of the results for scenarios 10 and 12 where the violations of normality are mild and for scenario 9 where the violation is severe.

**Conclusion**. For a continuous or a binary post-treatment confounder $Z$, the imputation-based strategy and the integration-based strategy both produce NIE and NDE estimates that are consistent with their respective benchmark values at the true value of $\rho_{10}$. This is true even when the structural part of the prediction model for $Z$ deviates from its data generation model. The results also remain robust when homoscedasticity is violated or when multivariate normality is violated to a minor degree. However, these strategies do not produce robust results when multivariate normality is severely violated. The simulation results provide important implications for data analysis. Since the true value of $\rho_{10}$ is unknown to the analyst, a sensitivity analysis must compare the initial estimates of NIE and NDE with the bounds for the new estimates obtained over the plausible range of $\rho_{10}$. By including strong predictors of $Z$ that help to increase the explained proportion of variation, the analyst may obtain relatively narrow bounds for the NIE and NDE estimates and thereby increasing the chance of arriving at a definite conclusion when initial results are indeed sensitive to the omission of post-treatment confounding.

## VI. NEWWS Application

In the initial analysis of the NEWWS-Riverside data, we select nine pretreatment covariates when fitting the propensity score model for the mediator under each treatment condition. The estimated effect size of NIE is -0.11 with a 95% confidence interval [-0.24, 0.01]; and that the estimated effect size of NDE is 0.13 with a 95% confidence interval [-0.09, 0.35]. Neither the NIE estimate nor the NDE estimate is statistically significant.

The above analysis did not adjust for potential post-treatment confounders that may bias the NIE and NDE estimates. We assess the sensitivity of the initial results to such omissions by applying the proposed strategies. We first evaluate the influence of a continuous post-treatment confounder—the amount of welfare received in the first year after the randomization. We specify a prediction model for this continuous $Z$ as a function of not just the nine pretreatment covariates that predicted the mediator but also two additional strong predictors of $Z$ (namely, welfare amount received in the pretreatment year and number of children in the household at baseline). To constrain the range of $\rho_{10}$, we further utilize a measure of duration of welfare dependence in the pretreatment year. Its partial correlations with $Z$ under the experimental and control conditions are -0.35 and -0.18, respectively. Applying Equation (2), the range of $\rho_{10}$ becomes [-0.86, 0.98], within which we choose 20 evenly spaced values. We then estimate NIE and NDE by adjusting for $Z$ along with the pretreatment covariates at each plausible value of $\rho_{10}$. To enhance the smoothness of the results, we conduct 200 imputations when implementing the imputation-based procedure.

Figures 6a and 6b display the estimation results obtained from the imputation-based analysis and the integration-based analysis, respectively. In each figure, the horizontal dashed line in the middle indicates the initial estimate of the NIE or NDE in effect size; the upper and lower dashed lines correspond to the initial 95% confidence interval of the effect size; each circle represents the adjusted estimate at a given value of $\rho_{10}$; and the corresponding vertical line represents the adjusted estimate of the 95% confidence interval. Over the range of the plausible values of $\rho_{10}$, we find that with adjustment for post-treatment welfare amount, the estimated effect size of NIE are bounded between -0.15 and -0.12 and that of NDE are bounded between 0.14 and 0.16. These are distinctly different from the initial estimate of the effect size of NIE (-0.11) and that of NDE (0.13). Apparently, omitting post-treatment welfare amount led to a positive bias in the initial NIE estimate and a negative bias in the initial NDE estimate. Comparing the adjusted confidence band to the initial confidence interval, we find that the statistical significance of NDE is insensitive to this omission; yet a small positive bias due to the omission has a consequence for the statistical significance of NIE. Hence, the latter is sensitive to the omission of post-treatment welfare amount at most plausible values of $\rho_{10}$.

We then evaluate the influence of a binary post-treatment confounder indicating whether an individual was on welfare continuously in the first year after the randomization. From both the imputation-based analysis and the integration-based analysis with adjustment for this binary $Z$, the bounds for the NIE estimates are (-0.15, -0.12) and those for the NDE estimates are (0.13, 0.16). Unsurprisingly, since this binary $Z$ (indicator for continuous welfare dependence) is closely associated with the continuous $Z$ (welfare amount), the sensitivity analysis arrives at a very similar conclusion.

## VII. Discussion

Some post-treatment confounders precede the focal mediator in the causal pathways while some others are concurrent to the focal mediator. This study is restricted to the former case. Qin, Deutsch, and Hong (2021) propose a sensitivity analysis strategy for causal mediation studies that involve two concurrent mediators that are not conditionally independent. Adjusting for selection bias associated with post-treatment confounders that precede the focal mediator has been a major challenge in causal mediation analysis that decomposes the average treatment effect into NIE and NDE. This is because a post-treatment covariate is observed under either the experimental condition or the control condition but not both. This study proposes a new approach to sensitivity analysis for assessing the consequences of omitting observed post-treatment confounders. The key is to obtain predicted values of the post-treatment covariate under the counterfactual control condition given an individual's observed post-treatment covariate value under the experimental condition as well as the observed pretreatment covariates. The analyst will then adjust for the observed and the predicted values of the post-treatment covariate through an RMPW analysis and obtain bounds for the NIE and NDE estimates. The analysis can be implemented through either imputation or integration over the conditional distribution of the post-treatment covariate. Unlike the linear structural modeling approach that requires correct specifications of not only the mediator model but also the outcome model, the weighting approach does not require the analyst to specify the response surface and therefore prevents bias induced by outcome model misspecification.

This study makes several important contributions to the literature on handling post-treatment confounding in causal mediation analysis. First, our new solution does not invoke additional identification assumptions beyond the standard sequential ignorability of the treatment assignment and of the mediator, which distinguishes our approach from alternative methods. Moreover, conditioning on not only the observed pretreatment covariates but also a potentially important post-treatment confounder increases the plausibility of mediator ignorability. Second, this new solution applies to both continuous and binary post-treatment covariates. Third, as we have shown through the simulations, the estimation results are robust even when the structural part of the prediction model for the post-treatment confounder deviates from its data generation model, when homoscedasticity is violated, or when multivariate normality is violated to a minor degree. Fourth, by obtaining the bounds for the NIE and NDE estimates and additionally by obtaining the confidence bands for these causal effects, we can assess whether adjustment for the post-treatment confounder would alter the initial conclusion in terms of not only the practical significance but also the statistical significance. And lastly, we underscore the practical value of including strong predictors of the post-treatment covariate in the prediction models as it will effectively make the bounds more informative for sensitivity analysis.

The simulation study reveals that the estimation results are not robust when multivariate normality is severely violated. In some cases, the analyst may overcome this limitation through appropriately transforming the post-treatment covariate. For example, if $Z$ follows a lognormal distribution, a log transformation will be a suitable solution. Future research may explore other potential solutions such as deriving the conditional distribution of $Z$ that follows a non-normal distribution or applying copula to the residuals obtained from the prediction models for $Z$.

## Acknowledgements

This study was supported by a grant from the National Science Foundation (SES 1659935) and a U.S. Department of Education Institute of Education Sciences (IES) Statistical and Research Methodology Grant (R305D120020). The authors would like to thank Li Cai, Donna Coffman, Trang Nguyen, Stephen Raudenbush, Geoff Wodtke, and participants at the University of California-Los Angeles Social Research Methodology - Human Development & Psychology Joint Brown Bag Speaker Series and at the University of Pittsburgh Department of Biostatistics Colloquium for their helpful comments.

**Supporting Information**

Web Appendices and Tables referenced in Sections V are available with this paper at the Biometrics website on Wiley Online Library.

Table 1. Simulation Results for Natural Indirect Effects

| Scenario | $\rho_{10}$ | True $NIE$ | $\widehat{NIE}_{w/oZ}$ | $\widehat{NIE}_{ora}$ | Imputation-Based RMPW Estimator | | | | Integration-Based RMPW Estimator | | | |
|---|---|---|---|---|---|---|---|---|---|---|---|---|
| | | | | | $\widehat{NIE}_{imp}$ bounds | $\widehat{NIE}_{imp}$ at $\rho_{10}$ | SE of $\widehat{NIE}_{imp}$ at $\rho_{10}$ | SD of $\widehat{NIE}_{imp}$ at $\rho_{10}$ | $\widehat{NIE}_{int}$ bounds | $\widehat{NIE}_{int}$ at $\rho_{10}$ | SE of $\widehat{NIE}_{int}$ at $\rho_{10}$ | SD of $\widehat{NIE}_{int}$ at $\rho_{10}$ |
| 1 | 0.5 | 0.352 | 0.310 | 0.351 | [0.314, 0.465] | 0.352 | 0.051 | 0.048 | [0.314, 0.465] | 0.352 | 0.047 | 0.048 |
| 2 | 0.0 | 0.387 | 0.306 | 0.385 | [0.311,0.458] | 0.384 | 0.056 | 0.051 | [0.311,0.458] | 0.385 | 0.049 | 0.050 |
| 3 | −0.5 | 0.425 | 0.311 | 0.425 | [0.316,0.463] | 0.426 | 0.059 | 0.057 | [0.316,0.463] | 0.426 | 0.054 | 0.056 |
| 4 | 0.5 | 0.352 | 0.208 | 0.348 | [0.313,0.449] | 0.348 | 0.167 | 0.161 | [0.313,0.449] | 0.348 | 0.152 | 0.161 |
| 5 | 0.0 | 0.387 | 0.308 | 0.383 | [0.316,0.452] | 0.384 | 0.178 | 0.165 | [0.316,0.452] | 0.385 | 0.160 | 0.165 |
| 6 | −0.5 | 0.425 | 0.311 | 0.418 | [0.316,0.452] | 0.417 | 0.186 | 0.186 | [0.316,0.452] | 0.419 | 0.172 | 0.188 |
| 7a | 0.8 | 0.333 | 0.296 | 0.336 | [0.320,0.489] | 0.337 | 0.049 | 0.049 | [0.320,0.489] | 0.337 | 0.047 | 0.049 |
| 7b | 0.5 | 0.333 | 0.293 | 0.333 | [0.318,0.378] | 0.333 | 0.049 | 0.046 | [0.318,0.378] | 0.333 | 0.047 | 0.046 |
| 8 | 0.4 | 0.090 | 0.032 | 0.087 | [0.071,0.141] | 0.090 | 0.058 | 0.056 | [0.071,0.141] | 0.090 | 0.055 | 0.056 |
| 9 | 0.2 | 0.375 | 0.364 | 0.374 | [0.345,0.374] | 0.354 | 0.047 | 0.036 | [0.345,0.374] | 0.354 | 0.046 | 0.047 |
| 10 | 0.0 | 0.260 | 0.248 | 0.258 | [0.248,0.263] | 0.256 | 0.088 | 0.082 | [0.248,0.263] | 0.256 | 0.082 | 0.082 |
| 11 | 0.5 | -0.423 | -0.352 | -0.427 | [-0.366,-0.450] | -0.427 | 0.048 | 0.047 | [-0.366,-0.450] | -0.427 | 0.046 | 0.047 |
| 12 | 0.0 | -0.365 | -0.272 | -0.360 | [-0.416,-0.305] | -0.361 | 0.053 | 0.048 | [-0.416,-0.305] | -0.361 | 0.048 | 0.047 |

Notes. Each scenario was simulated 1000 times. See the supporting information for details of data generation. This table lists the underlying true value of $\rho_{10}$. The true $NIE$ is calculated numerically by generating a dataset with a sample size of 5,000,000. $\widehat{NIE}_{w/oZ}$ is the mean of $NIE$ estimates across 1000 simulations without adjusting for $Z$ in the analysis; and $\widehat{NIE}_{ora}$ is the mean of $NIE$ estimates using the oracle estimator. For the imputation-based RMPW estimator (with 25 imputations), $\widehat{NIE}_{imp}$ bounds provide the means of the lower and upper bounds of $NIE$ estimates by varying $\rho_{10}$ in the range of $[-1, 1]$. At the true value of $\rho_{10}$, $\widehat{NIE}_{imp}$ at $\rho_{10}$ is the mean of $NIE$ estimates; SE of $\widehat{NIE}_{imp}$ at $\rho_{10}$ is the mean of standard errors of $NIE$ estimates; and SD of $\widehat{NIE}_{imp}$ at $\rho_{10}$ is the standard deviation of $NIE$ estimates across 1000 simulations. $\widehat{NIE}_{int}$, $\widehat{NIE}_{int}$ at $\rho_{10}$, SE of $\widehat{NIE}_{int}$ at $\rho_{10}$, and SD of $\widehat{NIE}_{int}$ at $\rho_{10}$ have analogous interpretations for the integration-based RMPW estimator.

Table 2. Simulation Results for Natural Direct Effects.

| Scenario | $\rho_{10}$ | True NDE | $\widehat{NDE}_{w/oZ}$ | $\widehat{NDE}_{ora}$ | Imputation-based RMPW Estimator | | | | Integration-based RMPW Estimator | | | |
|---|---|---|---|---|---|---|---|---|---|---|---|---|
| | | | | | $\widehat{NDE}_{imp}$ bounds | $\widehat{NDE}_{imp}$ at $\rho_{10}$ | SE of $\widehat{NDE}_{imp}$ at $\rho_{10}$ | SD of $\widehat{NDE}_{imp}$ at $\rho_{10}$ | $\widehat{NDE}_{int}$ bounds | $\widehat{NDE}_{int}$ at $\rho_{10}$ | SE of $\widehat{NDE}_{int}$ at $\rho_{10}$ | SD of $\widehat{NDE}_{int}$ at $\rho_{10}$ |
| 1 | 0.5 | 0.962 | 1.002 | 0.962 | [0.848, 0.998] | 0.961 | 0.039 | 0.030 | [0.848, 0.998] | 0.961 | 0.032 | 0.030 |
| 2 | 0.0 | 0.926 | 1.001 | 0.928 | [0.855,1.002] | 0.928 | 0.044 | 0.033 | [0.855,1.002] | 0.928 | 0.036 | 0.032 |
| 3 | −0.5 | 0.888 | 1.004 | 0.890 | [0.852,0.999] | 0.889 | 0.048 | 0.039 | [0.852,0.999] | 0.889 | 0.042 | 0.039 |
| 4 | 0.5 | 0.962 | 1.004 | 0.964 | [0.862,1.000] | 0.965 | 0.127 | 0.104 | [0.862,1.000] | 0.964 | 0.107 | 0.104 |
| 5 | 0.0 | 0.926 | 1.009 | 0.935 | [0.866,1.002] | 0.933 | 0.140 | 0.119 | [0.866,1.002] | 0.932 | 0.115 | 0.119 |
| 6 | −0.5 | 0.888 | 1.004 | 0.897 | [0.863,0.999] | 0.898 | 0.148 | 0.138 | [0.865,1.001] | 0.896 | 0.130 | 0.139 |
| 7a | 0.8 | 0.988 | 1.028 | 0.988 | [0.835,1.004] | 0.987 | 0.035 | 0.031 | [0.835,1.004] | 0.987 | 0.032 | 0.031 |
| 7b | 0.5 | 0.988 | 1.028 | 0.988 | [0.943,1.003] | 0.988 | 0.035 | 0.031 | [0.943,1.003] | 0.988 | 0.032 | 0.031 |
| 8 | 0.4 | 0.576 | 0.632 | 0.576 | [0.522,0.593] | 0.574 | 0.042 | 0.039 | [0.522,0.593] | 0.574 | 0.040 | 0.039 |
| 9 | 0.2 | 0.912 | 0.924 | 0.914 | [0.914,0.943] | 0.934 | 0.036 | 0.027 | [0.914,0.943] | 0.934 | 0.029 | 0.027 |
| 10 | 0.0 | 0.678 | 0.692 | 0.682 | [0.677,0.691] | 0.683 | 0.058 | 0.051 | [0.677,0.691] | 0.683 | 0.049 | 0.051 |
| 11 | 0.5 | 0.506 | 0.431 | 0.506 | [0.446,0.530] | 0.506 | 0.059 | 0.055 | [0.446,0.530] | 0.506 | 0.057 | 0.055 |
| 12 | 0.0 | 0.398 | 0.301 | 0.390 | [0.335,0.446] | 0.391 | 0.081 | 0.073 | [0.335,0.446] | 0.391 | 0.078 | 0.073 |

Notes. Each scenario was simulated 1000 times. See the supporting information for details of data generation. This table lists the underlying true value of $\rho_{10}$. The true $NDE$ is calculated numerically by generating a dataset of sample size of 5,000,000. $\widehat{NDE}_{w/oZ}$ is the mean of $NDE$ estimates across 1000 simulations without adjusting for $Z$ in the analysis; and $\widehat{NDE}_{ora}$ is the mean of $NDE$ estimates using the oracle estimator. For the imputation-based RMPW estimator (with 25 imputations), $\widehat{NDE}_{imp}$ bounds provide the means of the lower and upper bounds of $NDE$ estimates by varying $\rho_{10}$ in the range of $[-1, 1]$. at the true value of $\rho_{10}$, $\widehat{NDE}_{imp}$ at $\rho_{10}$ is the mean of $NDE$ estimates; SE of $\widehat{NDE}_{imp}$ at $\rho_{10}$ is the mean of standard errors of $NDE$ estimates; and SD of $\widehat{NDE}_{imp}$ at $\rho_{10}$ is the standard deviation of $NDE$ estimates across 1000 simulations. $\widehat{NDE}_{int}$, $\widehat{NDE}_{int}$ at $\rho_{10}$, SE of $\widehat{NDE}_{int}$ at $\rho_{10}$, and SD of $\widehat{NDE}_{int}$ at $\rho_{10}$ have analogous interpretations for the proposed integration-based RMPW estimator.

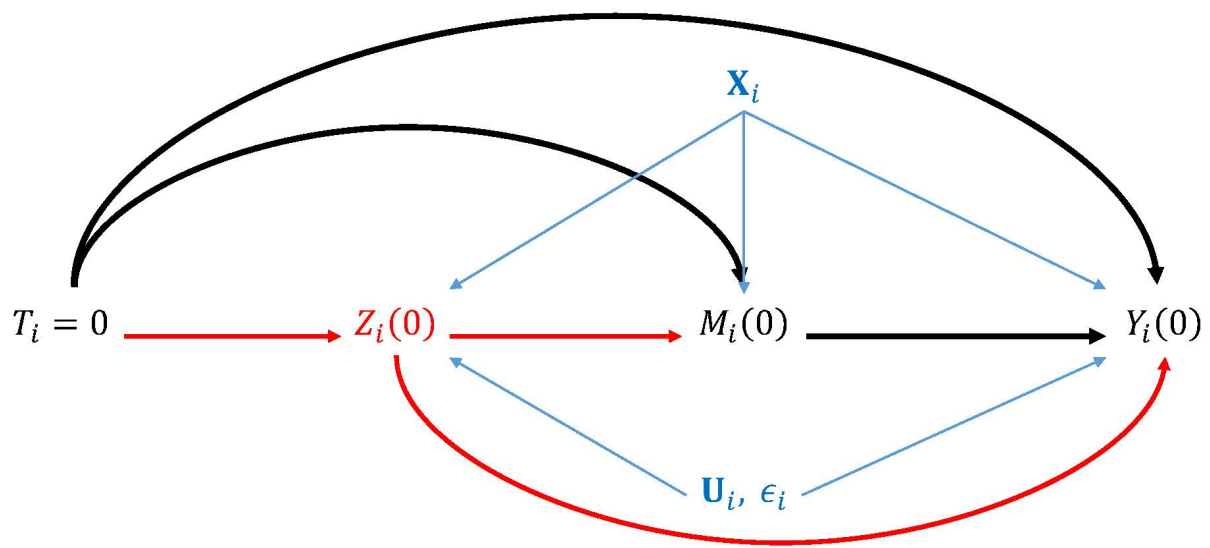

Figure 1a. Causal diagram for individual $i$ under the control condition

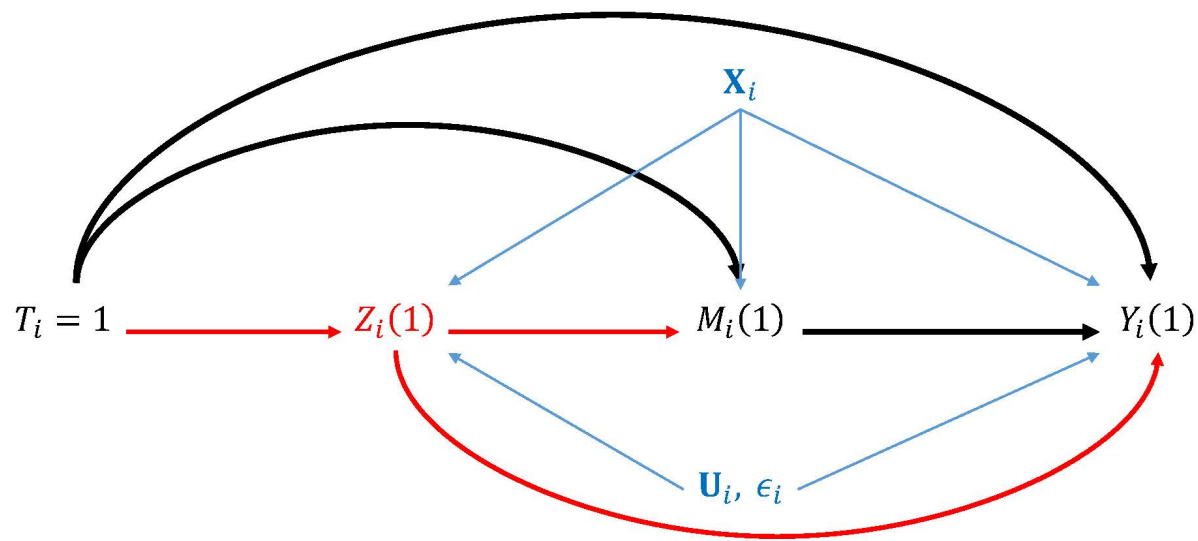

Figure 1b. Causal diagram for individual $i$ under the experimental condition

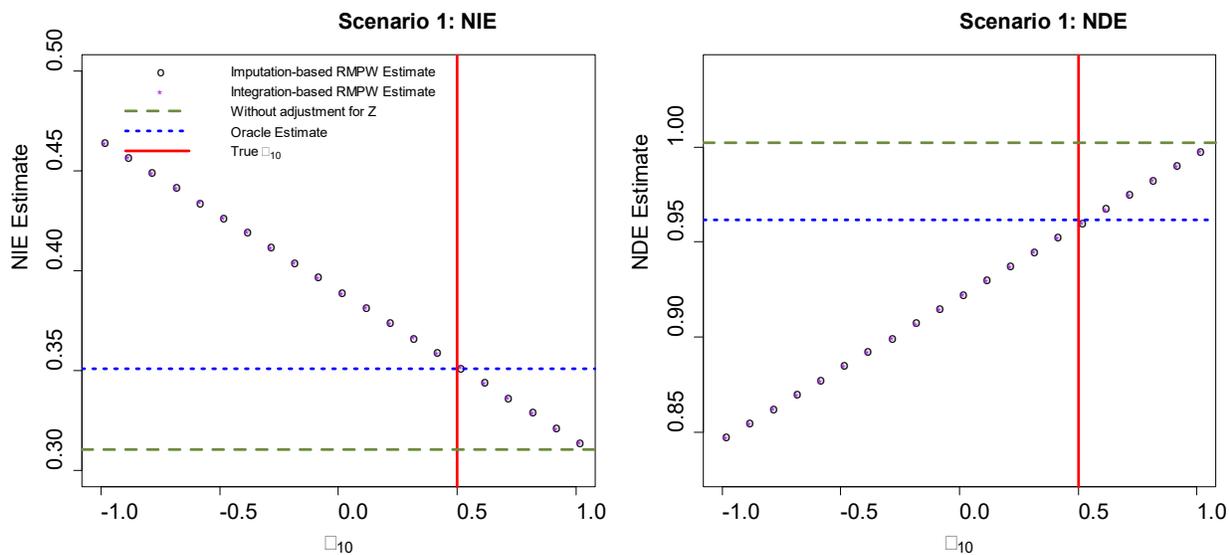

Figure 2. Simulation results for scenario 1

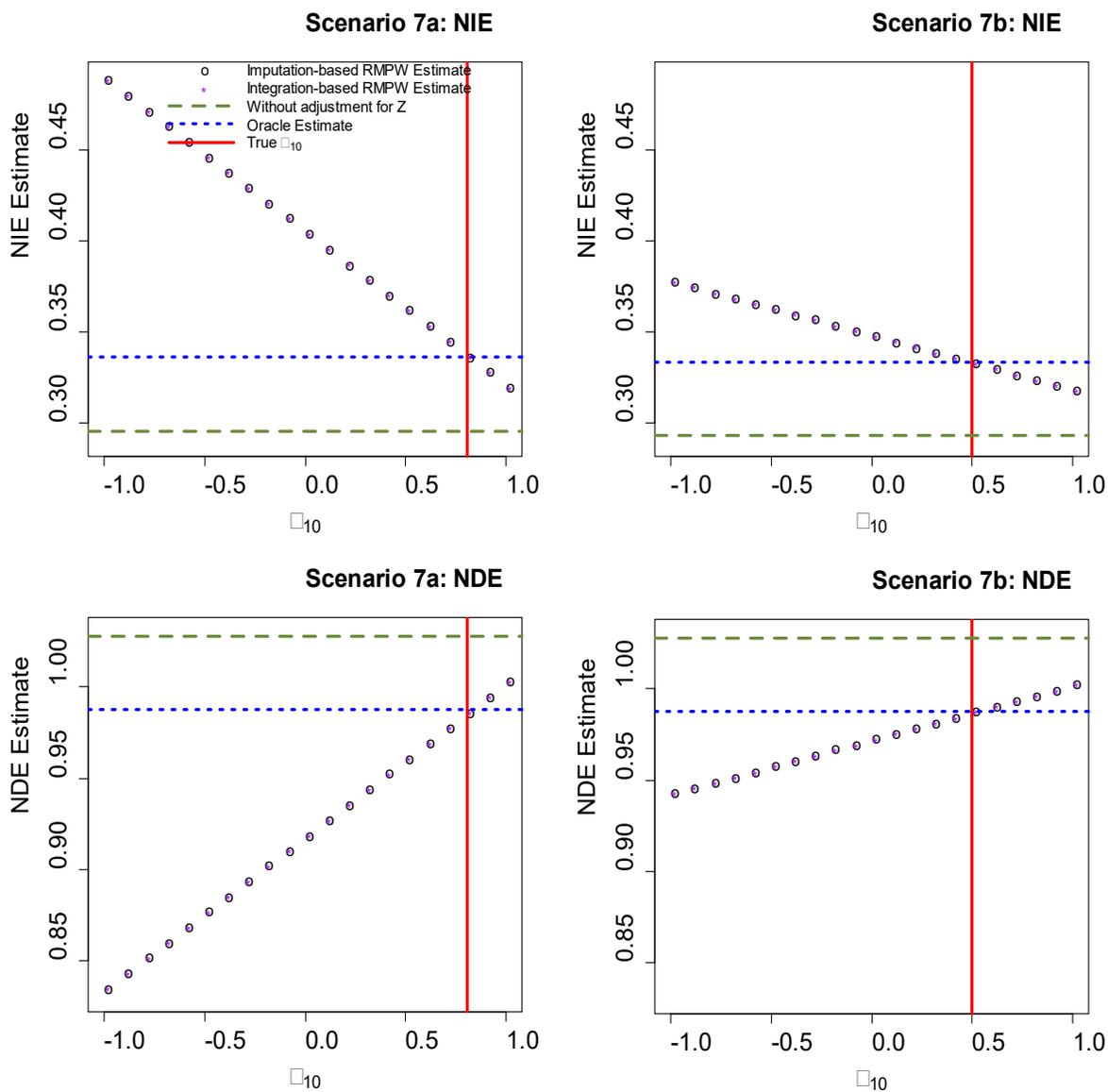

Figure 3. Simulation results comparing scenarios 7a and 7b

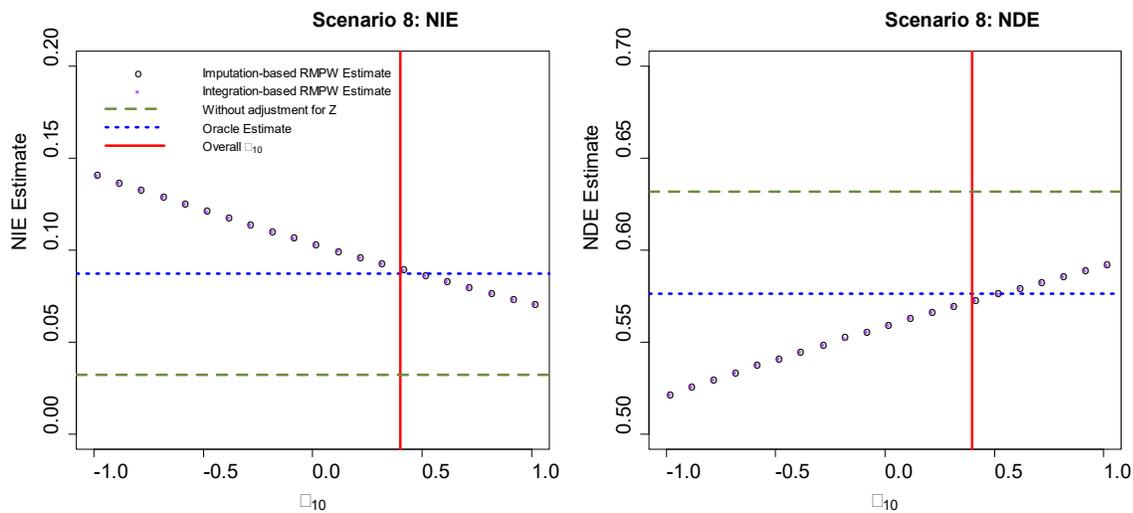

Figure 4. Simulation results for scenario 8

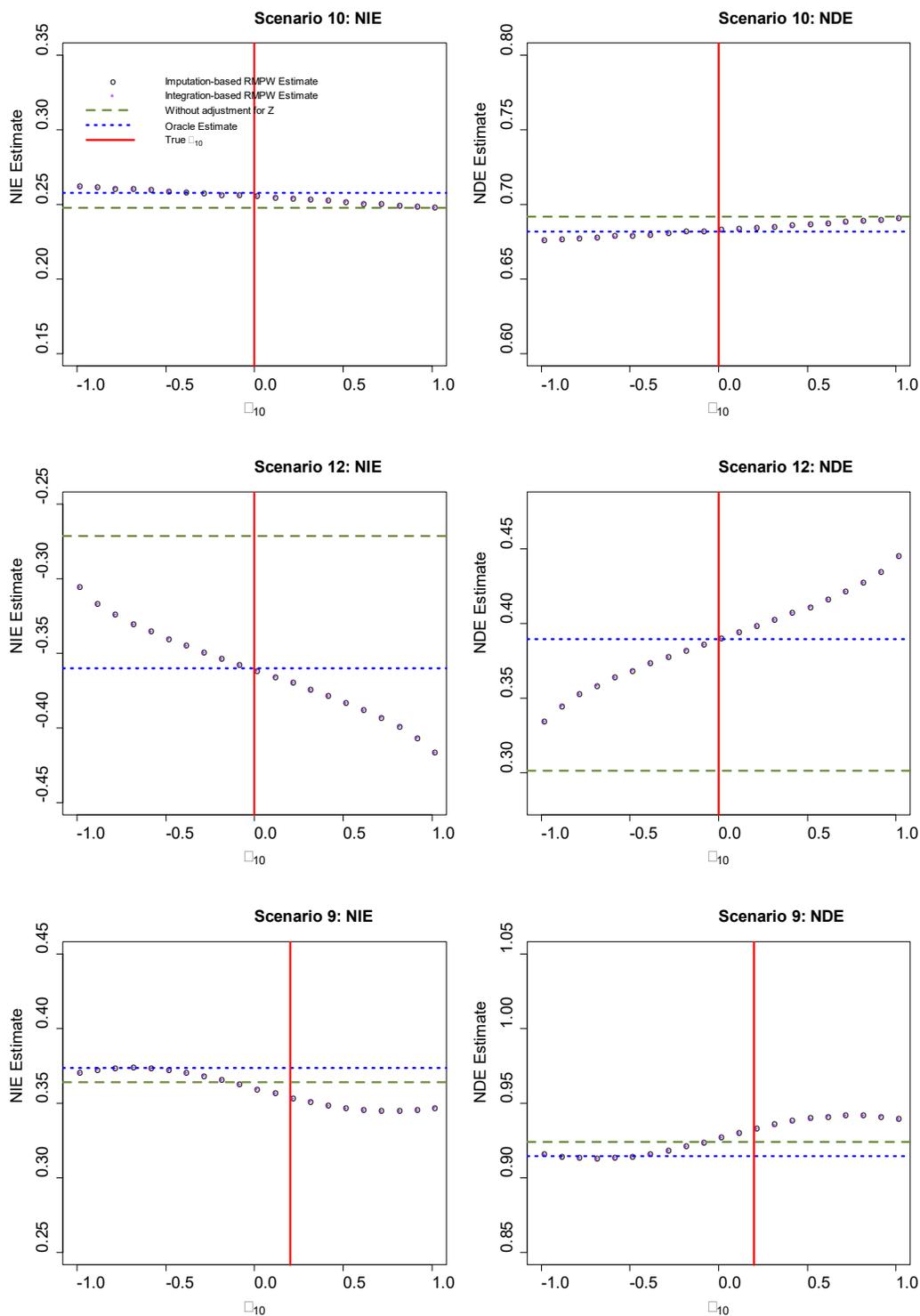

Figure 5. Simulation results when violations of multivariate normality are either mild (scenarios 10 and 12) or severe (scenario 9)

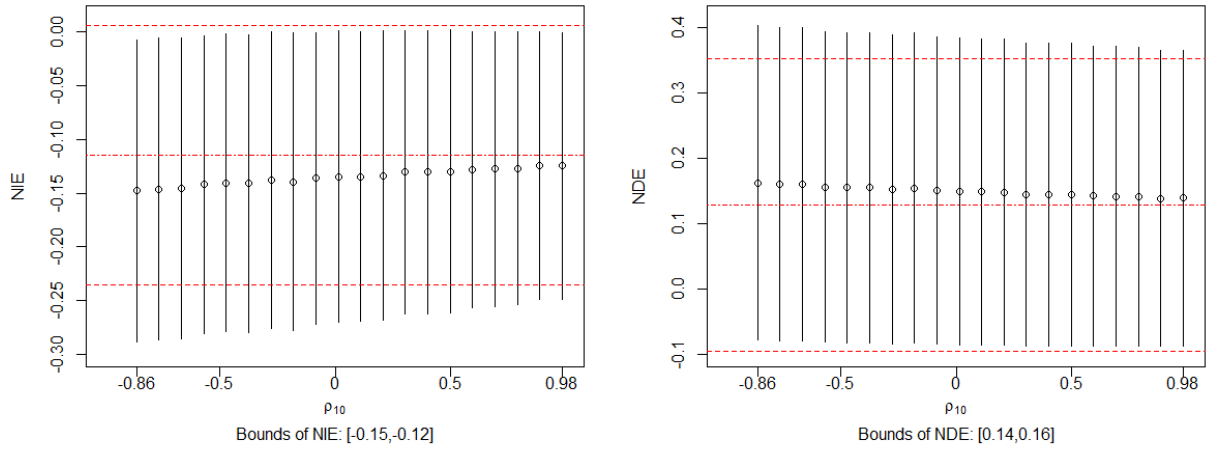

Figure 6a. Sensitivity of the initial estimate of the effect size of NIE and that of NDE to the omission of post-treatment welfare amount (imputation-based)

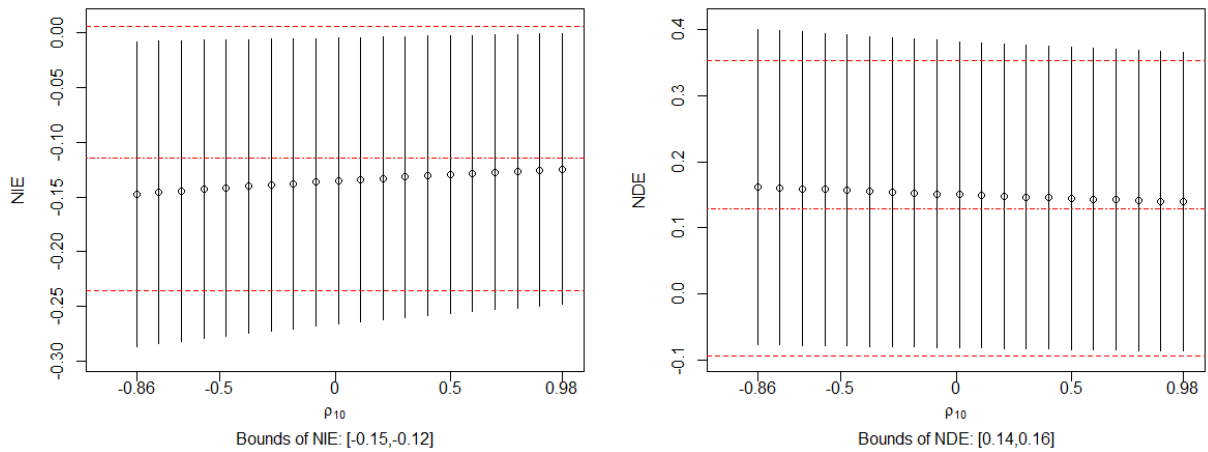

Figure 6b. Sensitivity of the initial estimate of the effect size of NIE and that of NDE to the omission of post-treatment welfare amount (integration-based)

**Supporting Information for "Post-Treatment Confounding in Causal Mediation Studies:**

**A Cutting-Edge Problem and A Novel Solution via Sensitivity Analysis"**

**by Guanglei Hong, Fan Yan, and Xu Qin**

**Web Appendix 1**

The variant of assumption (i) can be derived as follows. When assumption (i) holds, we have that, for $t, t' = 0,1$,

$$P(Y(t,m), M(t), M(t'), Z(t), Z(t') \mid T, \mathbf{X} = \mathbf{x})$$

$$= P(Y(t,m), M(t), M(t'), Z(t), Z(t') \mid \mathbf{X} = \mathbf{x}). \qquad (a1)$$

The left hand side of $(a1)$ is also equal to

$$P(Y(t,m), M(t), M(t') \mid T, Z(t), Z(t'), \mathbf{X} = \mathbf{x}) P(Z(t), Z(t') \mid T, \mathbf{X} = \mathbf{x})$$

$$= P(Y(t,m), M(t), M(t') \mid T, Z(t), Z(t'), \mathbf{X} = \mathbf{x}) P(Z(t), Z(t') \mid \mathbf{X} = \mathbf{x}).$$

The right hand side of (a1) is also equal to

$$P(Y(t,m), M(t), M(t') \mid Z(t), Z(t'), \mathbf{X} = \mathbf{x}) P(Z(t), Z(t') \mid \mathbf{X} = \mathbf{x}).$$

As the left hand side must be equal to the right hand side of $(a1)$, we obtain the following result:

$$P(Y(t,m), M(t), M(t') \mid T, Z(t), Z(t'), \mathbf{X} = \mathbf{x})$$

$$= P(Y(t,m), M(t), M(t') \mid Z(t), Z(t'), \mathbf{X} = \mathbf{x}).$$

Hence, assumption (i) implies the ignorability of the treatment assignment conditioning on $Z(t), Z(t')$, and $\mathbf{X}$.

**Web Appendix 2**

Here we prove the Theorem that obtains the conditional probability of the mediator under the counterfactual control condition for an individual assigned to the experimental condition when a post-treatment confounder exists. Let $h_{z'} = h_{iz'}(\mathbf{x}) = P(Z(0) = z' \mid T = 1, Z(1) = z, \mathbf{X} = \mathbf{x})$.

$$P(M(0) = m \mid T = 1, Z(1) = z, \mathbf{X} = \mathbf{x})$$

$$= \int P(M(0) = m, Z(0) = z' \mid T = 1, Z(1) = z, \mathbf{X} = \mathbf{x}) dz'$$

$$= \int P(M(0) = m \mid T = 1, Z(1) = z, Z(0) = z', \mathbf{X} = \mathbf{x}) h_{z'} dz'.$$

Under the variant of assumption (i), the above is equal to

$$\int P(M(0) = m \mid T = 0, Z(1) = z, Z(0) = z', \mathbf{X} = \mathbf{x}) h_{z'} dz'.$$

Then under the variant of assumption (ii), the above is equal to

$$\int P(M(0) = m \mid T = 0, Z(0) = z', \mathbf{X} = \mathbf{x}) h_{z'} dz'.$$

**Web Appendix 3**

The Lemma identifies the NIE and NDE when the treatment is randomized, which involves identifying $E\big[Y\big(1, M(0)\big)\big]$ through weighting. The proof is presented as follows:

$$E[Y(1, M(0))] = E[Y(1, M(0))|T = 1]$$

$$= E\{E[Y(1, M(0))|T = 1, Z(1) = z, \mathbf{X} = \mathbf{x}]|T = 1\}$$

$$= \iint \int y \times P(Y(1, m) = y|M(0) = m, T = 1, Z(1) = z, \mathbf{X} = \mathbf{x})$$

$$\times P(M(0) = m|T = 1, Z(1) = z, \mathbf{X} = \mathbf{x}) \times P(Z(1) = z|T = 1, \mathbf{X} = \mathbf{x})$$

$$\times P(\mathbf{X} = \mathbf{x}|T = 1) dy dm\, dz dx.$$

We now invoke assumption (ii) under which $M(0)$ and $M(1)$ are both independent of $Y(1, m)$ given $T = 1, Z(1) = z$, and $\mathbf{X} = \mathbf{x}$. The above is then equal to

$$= \iint \int y \times P(Y(1, m) = y|M(1) = m, T = 1, Z(1) = z, \mathbf{X} = \mathbf{x})$$

$$\times P(M(0) = m|T = 1, Z(1) = z, \mathbf{X} = \mathbf{x}) \times P(Z(1) = z|T = 1, \mathbf{X} = \mathbf{x})$$

$$\times P(\mathbf{X} = \mathbf{x}|T = 1) dy dm\, dz dx.$$

We then multiply and divide the above by $P(M(1) = m | T = 1, Z(1) = z, \mathbf{X} = \mathbf{x})$ and obtain the following:

$$\iint \iint \frac{P(M(0) = m | T = 1, Z(1) = z, \mathbf{X} = \mathbf{x})}{P(M(1) = m | T = 1, Z(1) = z, \mathbf{X} = \mathbf{x})} \times y$$

$$\times P(Y(1, m) = y | M(1) = m, T = 1, Z(1) = z, \mathbf{X} = \mathbf{x})$$

$$\times P(M(1) = m | T = 1, Z(1) = z, \mathbf{X} = \mathbf{x}) \times P(Z(1) = z | T = 1, \mathbf{X} = \mathbf{x})$$

$$\times P(\mathbf{X} = \mathbf{x} | T = 1) dy dm\, dz dx.$$

Given the theoretical result stated in the Theorem, the above is equal to

$$\iint \iint \frac{\int P(M(0) = m \mid T = 0, Z(0) = z', \mathbf{X} = \mathbf{x}) h_{z'} dz'}{P(M(1) = m | T = 1, Z(1) = z, \mathbf{X} = \mathbf{x})} \times y$$

$$\times P(Y(1, m) = y | M(1) = m, T = 1, Z(1) = z, \mathbf{X} = \mathbf{x})$$

$$\times P(M(1) = m | T = 1, Z(1) = z, \mathbf{X} = \mathbf{x}) \times P(Z(1) = z | T = 1, \mathbf{X} = \mathbf{x})$$

$$\times P(\mathbf{X} = \mathbf{x} | T = 1) dy dm\, dz dx$$

$$= E\left[\int W(z') Y h_{z'} dz' \mid T = 1\right],$$

where

$$W(z') = \frac{P(M(0) = m \mid T = 0, Z(0) = z', \mathbf{X} = \mathbf{x})}{P(M(1) = m | T = 1, Z(1) = z, \mathbf{X} = \mathbf{x})}.$$

When the treatment is randomized, $E[Y(t, M(t))] = E[Y(t, M(t)) | T = t] = E[Y | T = t]$ for $t = 0, 1$. NIE and NDE can then be identified as follows:

$$NIE = E[Y(1, M(1)) - Y(1, M(0))]$$

$$= E[Y(1, M(1))] - E[Y(1, M(0))]$$

$$= E[Y | T = 1] - E\left[\int W(z') Y h_{z'} dz' \mid T = 1\right].$$

Analogously,

$$NDE = E[Y(1, M(0)) - Y(0, M(0))]$$

$$= E[Y(1, M(0))] - E[Y(0, M(0))]$$

$$= E\left[\int W(z')Yh_{z'}dz' \,|T=1\right] - E[Y|T=0].$$

## Web Appendix 4

In this appendix we provide details of the data generating models for each of the twelve simulation scenarios. The conceptual relationships among the variables are represented in Figures 1a and 1b, in which Assumptions (i) and (ii) are satisfied.

We consider three different sample sizes in our simulations, $n=2,000$, $n=200$, and $n=700$, representing a large sample size, a small sample size, and a sample size similar to our application data, respectively. In each simulation scenario, we first generate a random treatment assignment variable $T_i \sim Bernoulli(\text{p})$. We consider a single pretreatment confounder $X_i$ in most simulation scenarios, except in scenario 8 where an additional binary pretreatment confounder $A_i$ is generated to examine the consequence of violating the homoscedasticity in the model for $Z$. The specifications for the sample size and for the distributions of $T_i$, $X_i$, and $A_i$ are presented in Table A1. These include continuous, binary, and categorical $X_i$.

We then generate $Z_i(t)$ for $t=0,1$ as a function of $X_i$ (and $A_i$ in scenario 8) and $r_{ti}$. For a continuous $Z$, we generate $Z_i(t) = \mu(t, X_i) + r_{ti}$ in general or $Z_i(t) = \mu(t, X_i, A_i) + r_{ti}$ in scenario 8. For a binary $Z$ , we first generate the latent continuous variable $Z_i^*(t) = \mu(t, X_i) + r_{ti}$ and then the binary $Z_i(0)$ by dichotomizing $Z_i^*(t)$ at 0, i.e., $Z_i(t) = I(Z_i^*(t) > 0)$.

Next, we generate a binary mediator $M_i(t)$ as follows:

$$P(M_i(t) = 1|Z_i(t), X_i, A_i) = \frac{\exp{(\alpha_{t,0} + \alpha_{t,Z}Z_i(t) + \alpha_{t,X}X_i + \alpha_{t,A}A_i)}}{1 + \exp{(\alpha_{t,0} + \alpha_{t,Z}Z_i(t) + \alpha_{t,X}X_i + \alpha_{t,A}A_i)}}, t = 0, 1,$$

where $\alpha_{t,A} = 0$ except in scenario 8. The specifications of the parameter values for the mediator models are presented in Table A2.

Finally, we generate a continuous outcome $Y_i(t)$ under treatment condition $t$ for $t = 0, 1$:

$Y_i(t) = \beta_{t,0} + \beta_{t,Z} Z_i(t) + \beta_{t,M} M_i(t) + \beta_{t,ZM} Z_i(0) \cdot M_i(0) + \beta_{t,X} X_i + \beta_{t,A} A_i + \beta_{t,r} r_{0i} + \eta_{ti}$,

where $\eta_{ti} \sim N(0, 0.1^2)$, and $\beta_{t,A} = 0$ except in scenario 8. The specifications of the parameter values for the outcome models are presented in Table A3. The observable post-treatment confounder, mediator, and outcome are $Z_i = T_i \cdot Z_i(1) + (1 - T_i) \cdot Z_i(0)$, $M_i = T_i \cdot M_i(1) + (1 - T_i) \cdot M_i(0)$ and $Y_i = T_i \cdot Y_i(1) + (1 - T_i) \cdot Y_i(0)$.

We now provide details of the data generating models for $Z$ and the purpose of each simulation scenario.

*Simulation Scenarios 1-6: When the model for a continuous $Z$ is correctly specified.*

In this set of simulations, we generate $Z_i(0)$ and $Z_i(1)$ according to the following joint distribution:

$$Z_i(0) = 0.5X_i + r_{0i};$$

$$Z_i(1) = 0.5 + X_i + r_{1i};$$

$$\begin{pmatrix} r_{0i} \\ r_{1i} \end{pmatrix} \sim N \left( \begin{pmatrix} 0 \\ 0 \end{pmatrix}, \quad \begin{pmatrix} 0.5 & \rho_{10}\sqrt{0.5} \\ \rho_{10}\sqrt{0.5} & 1 \end{pmatrix} \right).$$

Scenarios 1 to 3 considere a true value of $\rho_{10}$ being 0.5, 0, and -0.5, respectively when the sample size is 2,000; and in parallel, Scenarios 4 to 6 consider the same set of true values of $\rho_{10}$ when the sample size is 200. The prediction model for $Z$ in the data analysis is consistent with the data generation model.

*Simulation Scenario 7: When a greater proportion of variation in $Z$ is explained.*

In this simulation scenario, we evaluate the impact when the structural part of the prediction model for $Z$ deviates from the data generation model. We generate $Z_i(0)$ and $Z_i(1)$ according to the following joint distribution:

$$Z_i(0) = 0.5X_i + r_{0i};$$

$$Z_i(1) = 0.5 + X_i + r_{1i};$$

$$\begin{pmatrix} r_{0i} \\ r_{1i} \end{pmatrix} = \begin{pmatrix} L_i + r'_{0i} \\ 2L_i + r'_{1i} \end{pmatrix}, \text{ with } L_i \sim N(0, 0.5^2);$$

$$\begin{pmatrix} r'_{0i} \\ r'_{1i} \end{pmatrix} \sim \sim N\left( \begin{pmatrix} 0 \\ 0 \end{pmatrix}, \begin{pmatrix} 0.2 & 0.5 \cdot \sqrt{0.2 \cdot 0.4} \\ 0.5 \cdot \sqrt{0.2 \cdot 0.4} & 0.4 \end{pmatrix} \right).$$

Here $L$ is a predictor of $Z$ but not a predictor of $M$, and therefore, there is no need to adjust for $L$ for assumptions (i) and (ii) to hold. However, since $L$ is a strong predictor of $Z$, we expect that utilizing $L$ in predicting $Z$ may help improve the bounds on the effects estimates. We conduct a pair of simulations to evaluate this impact. In simulation 7a, we only use $X$ in the prediction models for $Z_i(0)$ and $Z_i(1)$. $X$ explains 12.2% and 15.2% of variations in $Z_i(0)$ and $Z_i(1)$, respectively, and the correlation of $r_{0i}$ and $r_{1i}$, denoted as $\rho_{10}$, equals 0.8. In simulation 7b, we use both $X$ and $L$ in the prediction models for $Z_i(0)$ and $Z_i(1)$. $X$ and $L$ jointly explain 61.0% and 75.8% of variations in $Z_i(0)$ and $Z_i(1)$, respectively, and the correlation of $r'_{0i}$ and $r'_{0i}$, denoted as $\rho'_{10}$, equals 0.5. The data generating models for other variables are the same as those in Scenario 1. The prediction models for $Z$ in data analysis match these two data generation models.

*Simulation Scenario 8: When homoscedasticity is violated in the model for a continuous $Z$.*

In this simulation, we examine the consequence of violating the homoscedasticity in the model for $Z$. We generate an additional binary covariate $A$, and let the variance and covariance structure of $\begin{pmatrix} r_{0i} \\ r_{1i} \end{pmatrix}$ vary across different subpopulations defined by $A$. We generate $Z_i(0)$ and $Z_i(1)$ according to the following joint distribution:

$$Z_i(0) = 0.5X_i + 0.1A_i + r_{0i};$$

$$Z_i(1) = 0.5 + X_i + 0.8A_i + r_{1i};$$

$$\begin{pmatrix} r_{0i} \\ r_{1i} \end{pmatrix} \sim N\left( \begin{pmatrix} 0 \\ 0 \end{pmatrix}, \begin{pmatrix} 0.8 & 0.6 \cdot \sqrt{0.8 \cdot 0.5} \\ 0.6 \cdot \sqrt{0.8 \cdot 0.5} & 0.5 \end{pmatrix} \right) \text{ when } A_i = 0;$$

$$\binom{r_{0i}}{r_{1i}} \sim N\left(\binom{0}{0}, \begin{pmatrix} 0.5 & 0.2 \cdot \sqrt{0.5} \\ 0.2 \cdot \sqrt{0.5} & 1 \end{pmatrix}\right) \text{ when } A_i = 1.$$

The overall correlation $\rho_{10}$ in the population (i.e., the correlation between $Z(0)$ and $Z(1)$ conditional on the structural model) is 0.4. In the data analysis, however, we incorrectly assume the variance and covariance structure to be homoscedastic in the prediction model for $Z$.

*Simulation Scenario 9: When the violation of the multivariate normality in the model for a continuous Z is severe.*

In this simulation, we assess the impact of a severe violation of the multivariate normality in the model for $Z$. Instead of generating $\binom{r_{0i}}{r_{1i}}$ from a bivariate normal distribution, we modify the simulation scenario 1 using a bivariate gamma distribution for $\binom{r_{0i}}{r_{1i}}$. We first generate $r_{0i}$ and $r_{1i}$ from a bivariate gamma distribution with a correlation 0.2, with a marginal distribution for $r_{0i} \sim Gamma(shape = 0.5, rate = 1)$ and a marginal distribution for $r_{1i} \sim Gamma(shape = 0.8, rate = 2)$; we then re-center $r_{0i}$ and $r_{1i}$ at 0 by subtracting their respective means. Other parts of the models are kept the same as those in Scenario 1. The data analysis incorrectly assumes normality.

*Simulation Scenario 10: When the violation of the multivariate normality in the model for a continuous Z is mild.*

In this simulation, we assess the impact of a mild violation of the multivariate normality in the model for $Z$. We generate $Z$ under a zero-inflated model where the proportions of 0 in the experimental group and the control group are the same as those in the real-data application. We consider a similar sample size to the real-data application, $n$=700, and a similar probability of treatment assignment. We generate $Z_i(0)$ and $Z_i(1)$ according to the following joint distribution:

$$Z_i(0) = I_i(0)(0.5 + 0.5X_i + r_{0i});$$

$$Z_i(1) = I_i(1)(0.7 + X_i + r_{1i});$$

$$\binom{r_{0i}}{r_{1i}} \sim N\left(\binom{0}{0}, \begin{pmatrix} 0.2^2 & 0 \\ 0 & 0.3^2 \end{pmatrix}\right);$$

$$I_i(0) \sim Bernoulli(0.93);$$

$$I_i(1) \sim Bernoulli(0.97).$$

The data analysis incorrectly assumes normality.

*Simulation Scenarios 11: When the model for a binary Z is correctly specified.*

In this simulation, we generate $Z_i(0)$ and $Z_i(1)$ according to the following joint distribution:

$$Z_i^*(0) = 0.2X_i + r_{0i}, \qquad Z_i(0) = I(Z_i^*(0) > 0);$$

$$Z_i^*(1) = 0.3 - 0.2X_i + r_{1i}, \qquad Z_i(1) = I(Z_i^*(1) > 0),$$

$$\binom{r_{0i}}{r_{1i}} \sim N\left(\binom{0}{0}, \begin{pmatrix} 1 & 0.5 \\ 0.5 & 1 \end{pmatrix}\right).$$

The data analysis is conducted by using the correct functional forms.

*Simulation Scenarios 12: When the model for a binary Z is incorrectly specified.*

In this simulation, we examine the case when the model for a binary $Z$ is incorrectly specified. Instead of using a bivariate normal model for $\binom{r_{0i}}{r_{1i}}$, we generate values of $\binom{r_{0i}}{r_{1i}}$ using logistic distributions. We generate $Z_i(0)$ and $Z_i(1)$ according to the following joint distribution:

$$Z_i^*(0) = 0.2X_i + r_{0i}, \qquad Z_i(0) = I(Z_i^*(0) > 0);$$

$$Z_i^*(1) = 0.3 - 0.2X_i + r_{1i}, \qquad Z_i(1) = I(Z_i^*(1) > 0);$$

where $r_{0i}$ and $r_{1i}$ are independently generated from a standard logistic distribution with a location parameter 0 and a scale parameter 1. The data analysis is conducted by incorrectly assuming normality.

**Web Table A1. Specifications of the sample size, and the models for T, X, and A**

| Scenario | Sample Size | $T_i$ | $X_i$ | $A_i$ |
|---|---|---|---|---|
| $1-3, 7, 9, 12$ | 2000 | $Bernoulli(0.5)$ | $N(0, 0.5^2)$ | — |
| $4-6$ | 200 | $Bernoulli(0.5)$ | $N(0, 0.5^2)$ | — |
| 8 | 2000 | $Bernoulli(0.5)$ | $N(0, 0.4^2)$ | $Bernoulli(0.5)$ |
| 10 | 700 | $Bernoulli(0.3)$ | $Bernoulli(0.5)$ | — |
| 11 | 2000 | $Bernoulli(0.5)$ | $P(X_i = x)$ $= \begin{cases} 0.25, x = -1; \\ 0.25, x = 0; \\ 0.20, x = 1; \\ 0.30, x = 1.5 \end{cases}$ | — |

**Web Table A2. Specifications of the parameter values for the mediator models**

| Scenario | $\alpha_{0,0}$ | $\alpha_{0,Z}$ | $\alpha_{0,X}$ | $\alpha_{0,A}$ |
|---|---|---|---|---|
| $1-7,9,10$ | 0 | 1 | 0.7 | 0 |
| 8 | 0 | 0.5 | 0.4 | 1 |
| 11,12 | 0 | 1.2 | 0.7 | 0 |
|  |  |  |  |  |
| Scenario | $\alpha_{1,0}$ | $\alpha_{1,Z}$ | $\alpha_{1,X}$ | $\alpha_{1,A}$ |
| $1-7,9,10$ | 0.5 | 0.5 | 1.5 | 0 |
| 8 | 0.3 | 1 | 1 | $-1$ |
| 11,12 | 0.3 | $-0.7$ | 0.3 | 0 |

**Web Table A3. Specifications of the parameter values for the outcome models**

| Scenario | $\beta_{0,0}$ | $\beta_{0,Z}$ | $\beta_{0,M}$ | $\beta_{0,ZM}$ | $\beta_{0,X}$ | $\beta_{0,A}$ | $\beta_{0,r}$ |
|---|---|---|---|---|---|---|---|
| $1-7,9,10$ | 0 | 0.5 | 1 | $-0.5$ | 0.4 | 0 | 0.2 |
| 8 | 0 | 0.5 | 1 | $-0.5$ | 0.4 | 0.7 | 0.2 |
| 11,12 | 0 | 1 | 1 | $-1$ | 0.4 | 0 | 0.8 |
|  |  |  |  |  |  |  |  |
| Scenario | $\beta_{1,0}$ | $\beta_{1,Z}$ | $\beta_{1,M}$ | $\beta_{1,ZM}$ | $\beta_{1,X}$ | $\beta_{1,A}$ | $\beta_{1,r}$ |
| $1-7,9,10$ | 0.7 | $-1$ | 2 | 0.5 | 0.5 | 0 | 0.4 |
| 8 | 0.7 | $-1$ | 2 | 0.5 | 0.5 | 0.3 | 0.4 |
| 11,12 | 0.7 | $-1.5$ | 1.5 | 1 | 0.5 | 0 | 1 |